\documentclass[preprint, floatfix]{revtex4}

\usepackage[final]{epsfig}
\usepackage{amsmath}
\usepackage{parskip}

\begin{document}

\title{A model for the spacetime evolution of heavy-ion collisions at RHIC}

\author{Thorsten Renk}

\pacs{25.75.-q}
\preprint{DUKE-TH-04-263}

\affiliation{Department of Physics, Duke University, PO Box 90305,  Durham, NC 27708 , USA}

\begin{abstract}

We investigate the space-time evolution of ultrarelativistic Au-Au
collisions at full RHIC energy using a schematic model of the expansion. 
Assuming a thermally equilibrated system, we can adjust the essential
scale parameters of this model such that the measured transverse momentum spectra and
Hanbury-Brown Twiss (HBT) correlation parameters are well described. We find that the experimental
data strongly constrain the dynamics of the evolution of the emission source although hadronic observables
for the most part reflect the final breakup of the system.

\end{abstract}

\maketitle

\section{Introduction}
\label{sec_introduction}

Numerical simulations of finite temperature Quantum Chromodynamics (QCD) on discrete 
space-time lattices suggest that the theory exhibits a transition to a new state of
matter, the quark-gluon plasma (QGP), at temperatures of the order of 170 MeV \cite{Lattice}.
Experimentally, hadronic matter at this temperature can only be created for a short
time in ultrarealtivistic heavy-ion collisions, and currently there are ongoing
efforts from both experiment and theory to find conclusive evidence for the
creation of the QGP and ultimately to study its properties. A wealth of data
on hadronic single particle distributions and two particle correlations has been
assembled so far.

In the interpretation of experimental data of ultrarelativistic heavy-ion collisions, however, one is 
often faced with the challenge to disentangle signals of new physics indicating 
the production of a QGP from known hadronic effects. This question can only be reliably
addressed if the space-time evolution of the fireball created in
the collision is sufficiently known. In this note, we present an attempt to determine the
evolution of the bulk hadronic matter by fitting the essential scales of a
schematic model for the fireball expansion dynamics to the measured hadronic data. 
The resulting scenario can then be used
in the calculation of other observables (not used in the fit) which are 
sensitive to the bulk matter expansion, such as
the emission of dileptons and photons or jet quenching, thus reducing or eliminating
the inherent ambiguity
in the interpretation mentioned above.

\section{The model}

The main assumption for the model is that an equilibrated system is formed
a short time $\tau_0$ after the onset of the collision. Furthermore, we assume that this
thermal fireball subsequently expands isentropically until the mean free path of particles exceeds
(at a timescale $\tau_f$) the dimensions of the system and particles 
move without significant interaction to the detector.
In addition to this final breakup (freeze-out) of the fireball, particles are emitted 
throughout the expansion period whenever they cross
the boundary of the thermalized fireball matter.

For simplicity we restrict the discussion to a system exhibiting radial symmetry around the beam (z)-axis
corresponding to a central ($b=0$) collision.
For the entropy density at a
given proper time we make the ansatz 
\begin{equation}
s(\tau, \eta_s, r) = N R(r,\tau) \cdot H(\eta_s, \tau)
\end{equation}
with $\tau $ the proper time as measured in a frame co-moving
with a given volume element and $\eta_s = \frac{1}{2}\ln (\frac{t+z}{t-z})$ the spacetime
rapidity and $R(r, \tau), H(\eta_s, \tau)$ two functions describing the shape of the distribution
and $N$ a normalization factor.
We use Woods-Saxon distributions
\begin{equation}
\begin{split}
&R(r, \tau) = 1/\left(1 + \exp\left[\frac{r - R_c(\tau)}{d_{\text{ws}}}\right]\right)
\\ & 
H(\eta_s, \tau) = 1/\left(1 + \exp\left[\frac{\eta_s - H_c(\tau)}{\eta_{\text{ws}}}\right]\right).
\end{split}
\end{equation}
to describe the shapes for a given $\tau$. Thus, the ingredients of the model are the 
skin thickness parameters $d_{\text{ws}}$ and $\eta_{\text{ws}}$
and the para\-me\-tri\-zations of the expansion of the spatial extensions $R_c(\tau), H_c(\tau)$ 
as a function of proper time.  For simplicity, we assume for the moment a radially non-relativistic 
expansion and constant acceleration, therefore we find
$R_c(\tau) = R_0 + \frac{a_\perp}{2} \tau^2$. $H_c(\tau)$ is obtained
by integrating forward in $\tau$ a trajectory originating from the collision center which is characterized
by a rapidity  $\eta_c(\tau) = \eta_0 + a_\eta \tau$
with $\eta_c = \text{atanh } v_z^c$ the longitudinal expansion velocity for that 
trajectory. Since the relation between proper time as measured in the co-moving frame 
and lab time is determined by the rapidity at a given time, the resulting integral is
in general non-trivial and solved numerically (see \cite{Synopsis} for details).
$R_0$
is determined in overlap calculations using Glauber theory, the initial size of the rapidity interval occupied
by the fireball matter. $\eta_0$ is a free parameter and we choose to use the transverse velocity  $v_\perp^f = a_\perp \tau_f$ and
rapidity at decoupling proper time $\eta^f = \eta_0 + a_\eta \tau_f$ as parameters.
Thus, specifying $\eta_0, \eta_f, v_\perp^f$ and $\tau_f$ sets the essential scales of the spacetime
evolution and $d_{\text{ws}}$ and $\eta_{\text{ws}}$ specify the detailed distribution of entropy density.
For simplicity, we do not discuss a (possible) time dependence of the shapes (e.g. parametrized
by  $d_{\text{ws}}(\tau)$ and $\eta_{\text{ws}}(\tau)$) at this point.

We require that the parameter set $(\eta_f, \eta_{\text{ws}})$ reproduces the experimentally observed rapidity
distribution of particles \cite{Brahms-dN-deta} but we do not make any assumptions about the initial
rapidity interval characterized by $\eta_0$. This allows for the possibility of accelerated longitudinal expansion
and implies in general $\eta \neq \eta_s$. Here, $\eta = \frac{1}{2} \ln\frac{p_0 + p_z}{p_0 - p_z}$
denotes the longitudinal rapidity of a volume element moving with 
momentum $p^\mu$. We require that the longitudinal flow profile is
such that an initially homogeneous distribution remains homogeneous.
For a general accelerated motion, there is no simple analytical expression
for the spacetime position of sheets of given $\tau$ or the flow
profile corresponding to $\tau = \sqrt{t^2 - z^2}$ and $\eta = \eta_s$ 
valid for the non-accelerated case. 
In \cite{Synopsis} however, we 
have investigated such an expansion pattern with a constant acceleration and 
argued that for rapidities $<4$ the acceleration leads to an approximately 
linear relation $\eta \approx \zeta \eta_s$ at given $\tau$ and that sheets of
constant proper time are approximately hyperbolae as in the non-accelerated case.
This mismatch between $\eta$ and $\eta_s$ leads to additional Lorentz contraction
factors in volume integrals at given $\tau$, hence
the longitudinal
extension of matter on a sheet of given $\tau$ in
the interval $ -\eta_s^\text{front} < \eta_s < \eta_s^\text{front}$ must then be
calculated as
\begin{equation}
L(\tau) \approx 2 \tau \frac{\text{sinh }(\zeta -1) \eta_s^\text{front}}{(\zeta -1)}.
\end{equation}
In the following, we use these results in our computations whenever
$\eta_f \neq \eta_0$. For $\eta_f = \eta_0$, the model reduces to the well-known
expressions of the Bjorken expansion scenario, e.g. $L(\tau) = 2 \eta_0 \tau$
with $\tau = \sqrt{t^2 - z^2}$.

For transverse flow we assume a linear relation between radius $r$ and
transverse rapidity $\rho = \text{atanh } v_\perp(\tau) =  r/R_c(\tau) \cdot \rho_c(\tau)$
with $\rho_c(\tau) = \text{atanh } a_\perp \tau$.
For the net baryon density inside the fireball matter we assume a transverse distribution
(apart from a normalization factor) given by $R(r,\tau)$, but its longitudinal distribution
we parametrize such as to describe the measured data \cite{Brahms}.

We proceed by specifying the equation of state (EoS) of the thermalized matter. In the QGP phase,
we use an equation of state based on a quasiparticle interpretation of lattice QCD data
(see \cite{Quasiparticles}). In the hadronic phase, we adopt the picture
of subsequent chemical freeze-out (at the transition temperature $T_C$) and thermal freeze-out (at 
breakup temperature $T_F$)
and consequently  use a resonance gas EoS 
which depends on the local net baryon density.
The reason for this is that a finite ba\-ryo\-chemical potential $\mu_B$ leads to an increased number of
heavy resonances at the phase transition point, and decay pions from these resonance decays
 lead in turn to a finite pion chemical potential $\mu_\pi$ in the late evolution phases which implies
an overpopulation of pion phase space and faster cooling as compared to
a scenario in chemical equilibrium. We calculate the pion chemical potential
as a function of the local baryon and entropy density using the statistical hadronization framework 
outlined in \cite{Hadrochemistry}. We find $\mu_\pi$ to be small ($\mathcal{O}(30\text{ MeV})$) in the
midrapidity region at RHIC, however the corrections to the chemically equilibrated case are
important.

With the help of the EoS, we can find the local temperature $T(\eta_s, r, \tau)$ of a volume element
from its entropy density $s(\eta_s, r, \tau)$ and net baryon density $\rho_B(\eta_s,r,\tau)$.
We calculate particle emission throughout the whole lifetime of the fireball
by selecting a freeze-out temperature $T_f$, finding the hypersurface characterized by
$T(\eta_s, r, \tau) = T_f$ and evaluating the Cooper-Frye
formula
\begin{equation}
E \frac{d^3N}{d^3p} =\frac{g}{(2\pi)^3} \int d\sigma_\mu p^\mu
\exp\left[\frac{p^\mu u_\mu - \mu_i}{T_f}\right] = d^4 x S(x,p)
\end{equation}
with $p^\mu$ the momentum of the emitted particle and $g$ its 
degeneracy factor.
Note that the factor $d\sigma_\mu p^\mu$ contains the spacetime rapidity
$\eta_s$ and the factor $p^\mu u_\mu$ the rapidity $\eta$. Since these are
in general not the same in our model, the analytic expressions
valid for a boost-invariant scenario \cite{ThermalPhenomenology} 
do not apply. 
\begin{figure*}[!htb]
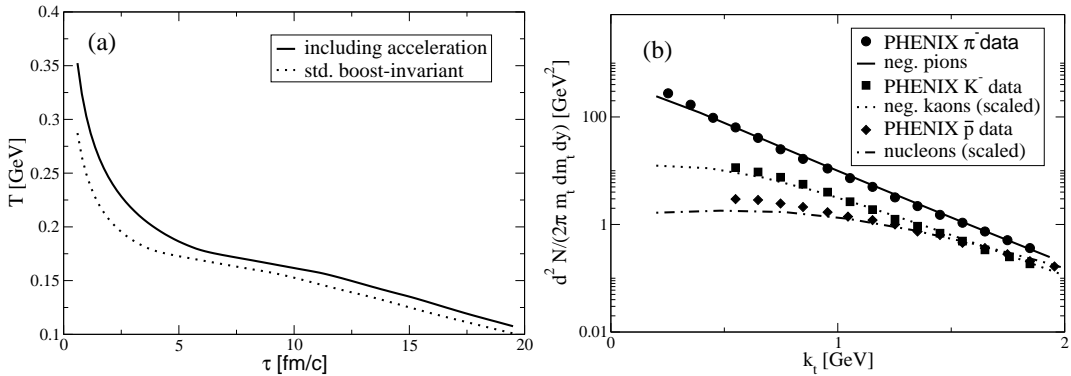

\begin{center}
\epsfig{file=cooling_curve.eps, width = 7.0cm}
\epsfig{file=spectra.eps, width=7.0cm}
\end{center}
\caption{\label{F-Spec}Left panel: Time evolution of the average fireball temperature
for the fitted set of parameters (solid line) and a model assuming boost-invariant longitudinal
expansion (dotted). 
Right panel: Measured transverse momentum spectra
for $\pi^-$ (circles), $K^-$ (squares) and $\overline{p}$ (diamonds) as compared to the
model results. }
\end{figure*}

In our case, the parameter $\tau_f$ denotes the freeze-out time for the
last volume element to reach the temperature $T_f$.
Volume elements freeze out throughout the whole evolution
time (and are boosted with a time- and position dependent flow velocity) whenever they
cross the Cooper-Frye hypersurface. Therefore the transverse expansion parameter $v_\perp^f$ also does
not necessarily reflect the typical transverse velocity of a volume element since
freeze-out may occur earlier than $\tau_f$  and not at the radius $r = R(\tau)$. 

Using this emission function, we calculate the HBT parameters as \cite{HBTReport, HBTBoris}
\begin{equation}
R_{\text{side}}^2 = \langle \tilde{y}^2 \rangle \quad R_{\text{out}}^2 =\langle (\tilde{x} -\beta_\perp \tilde{t})^2 
\rangle \quad R_{\text{long}} = \langle \tilde{z}^2 \rangle
\end{equation}
with $\tilde{x}_\mu = x_\mu -\langle x_\mu \rangle$ and
\begin{equation}
\langle f(x)\rangle(K) = \frac{\int d^4 x f(x) S(x,K)}{\int d^4x S(x,K)}
\end{equation}
In order to parametrize the spacetime evolution of a central 200 AGeV Au-Au
collision at RHIC we fit the remaining set of parameters $d_{\text{ws}}, \eta_{0},
T_F$ and $v_\perp^f$ to the experimentally obtained single particle
transverse momentum spectra \cite{Phenix-HBT} and HBT parameters \cite{Phenix-Spectra}.
For the description of the HBT correlation parameters which are measured for
30\% central collisions, we scale down the entropy content and initial overlap
radius guided by overlap calculations and neglect the angular asymmetry.

\section{Results and discussion}

\begin{figure*}[!htb]
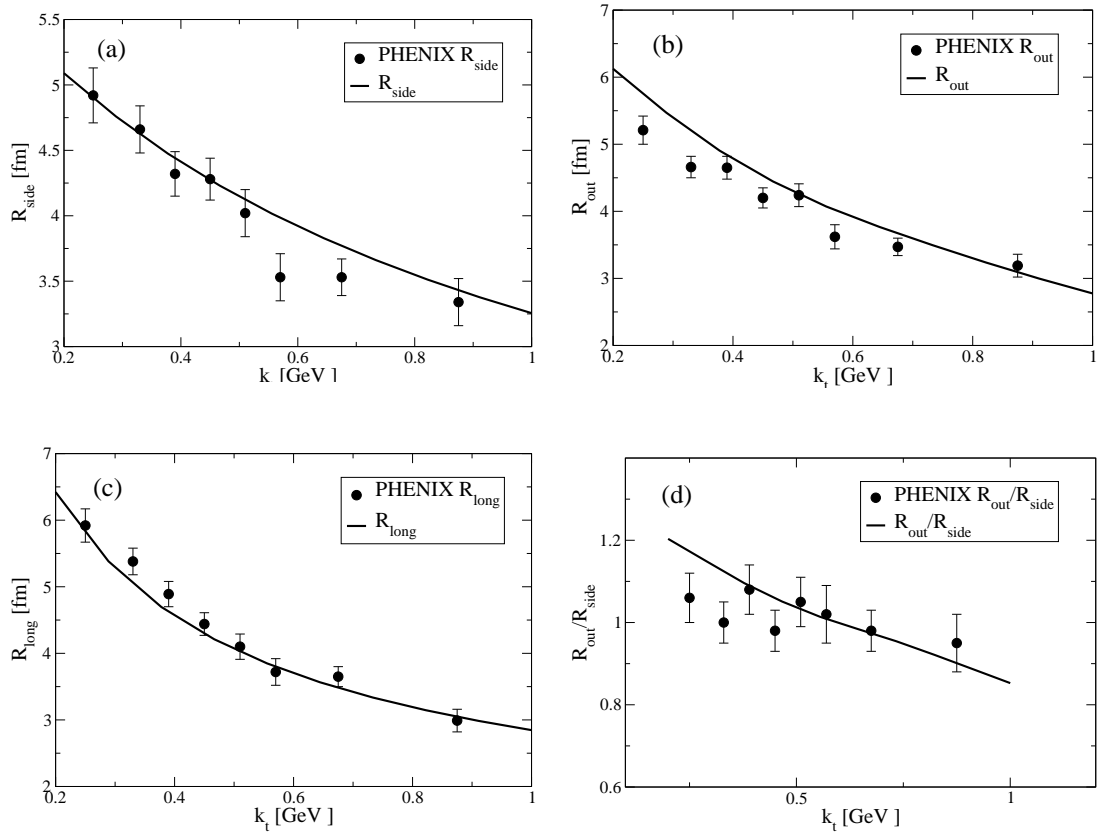

\begin{center}
\vspace{0.35cm}
\epsfig{file=R_side.eps, width = 7.0cm} \hspace{0.2cm}
\epsfig{file=R_out.eps, width=7.0cm}\\ \vspace{0.7cm}
\epsfig{file=R_long.eps, width = 7.0cm}  \hspace{0.2cm}
\epsfig{file=R_out_side_ratio.eps, width=7.0cm}
\end{center}
\caption{\label{F-HBT}The HBT correlation parameters $R_{\text{side}}, R_{\text{out}}, R_{\text{long}}$
and the ratio $R_{\text{out}}/R_{\text{side}}$ in the model calculation as compared to PHENIX
data \cite{Phenix-HBT}.}
\end{figure*}

We find that the choice of parameters $T_f$ = 110 MeV,  $d_{\text{ws}} <$  1.0 fm, 
$\eta_0$ = 1.8 and $v_\perp^f = 0.67$ is able to give a good description of the data.
The resulting average cooling curve (computed by averaging the entropy density
over the fireball volume at a given proper time and determining the corresponding temperature, where
we define the fireball volume at given $\tau$ as the 3-volume bounded by the Cooper-Frye surface) 
and the transverse momentum spectra
for $\pi^-, K^-$ and nucleons are shown in Fig.~\ref{F-Spec}, the HBT correlation parameters
in Fig.~\ref{F-HBT}.

The fit apparently misses the low $p_t$ part of the transverse momentum spectra for
the heavier particles
but describes the $\pi^+\pi^+$ correlation radii well, with the exception of 
the low $p_t$ part of $R_{\text{out}}$ where the calculation lies somewhat above the data.

\begin{figure*}[!htb]
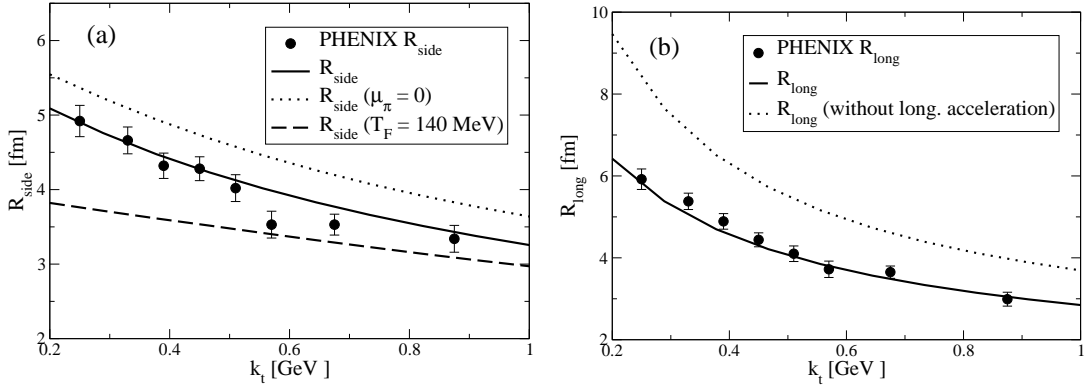

\begin{center}
\epsfig{file=R_side_wrong.eps, width = 7.0cm}\hspace{0.2cm}
\epsfig{file=R_long_wrong.eps, width= 7.0cm}

\end{center}
\caption{\label{F-HBTwrong}Left panel: $R_{\text{side}}$ in a scenario with vanishing pion
chemical potential $\mu_\pi$ (dotted line) and a difference scenario assuming $T_f = 140$ MeV (dashed line)
as compared to the standard calculation (solid) Right panel: $R_{\text{long}}$ in a scenario with
$T_f$ = 110 MeV but assuming no longitudinal acceleration (dotted) compared with
the standard scenario (solid). Note that the
transverse momentum spectra are well described by all of the scenarios shown in this
figure. }
\end{figure*}

The rather steep falloff of $R_{\text{side}}$, $R_{\text{out}}$ as a function of $k_t$ favours a large amount of
flow. However, demanding simultaneous agreement with the slope of the momentum spectra implies
that large transverse flow has to be accompanied by low freeze-out temperatures. Therefore, the volume
at freeze-out has to be large in order to reach small entropy densities. The inclusion of
a finite pion chemical potential (originating from resonance decays 
in chemical non-equilibrium)
is crucial to
reduce the volume corresponding to a given average temperature. Since the radial
expansion is rather constrained by the normalization of the HBT parameters, a large volume
implies sizeable longitudinal extension and hence a long-lived system. Such a long
lifetime in combination with a standard boost-invariant longitudinal expansion leads
to large values of $R_{\text{long}}$ which are clearly incompatible with the data. A sizeable
longitudinal compression and re-expansion however leads to a reasonable description
of $R_{\text{long}}$ as well.

It is not possible to identify a single feature of the model as being responsible for
the good quality of the fit but one can illustrate certain
trends. Neglecting the additional cooling caused by a finite value of $\mu_\pi$, it is still possible to
describe the spectra and the slope of the HBT data well for a low freeze-out 
temperature, however a larger volume is needed to get to this temperature 
and therefore the normalization of the correlation radii is systematically above the data
(see Fig.~\ref{F-HBTwrong}, left panel).

On the other hand, retaining the idea of a boost-invariant scenario without longitudinal
acceleration implies a short lifetime of the system as apparent from the estimate
$
\label{E-Sinyukov}
R_{\text{long}} = \tau_f(T_f/m_t)^{1/2}
$
based on such an expansion pattern. Such a short lifetime can be achieved
by a large value of $T_f$. The transverse expansion velocity can then be adjusted
to fit the $m_t$ spectra, but the resulting solution is characterized by small flow velocities
which leave both normalization and falloff of $R_{\text{side}}$ in disagreement with the data 
(see Fig.~\ref{F-HBTwrong}, left panel). Retaining a low freeze-out temperature in
a boost-invariant scenario allows to describe the transverse spectra and correlations well
but significantly overpredicts $R_{\text{long}}$ (see Fig.~\ref{F-HBTwrong}, right panel).

Concluding, we find that the $m_t$ spectra and HBT parameters measured at RHIC can
be simultaneously described assuming a scenario with small freeze-out temperature $T_F\approx 110$ MeV
and a sizeable initial longitudinal compression and re-expansion. The data strongly constrain
alternative scenarios. An upcoming publication will
investigate the dependence of the results on the different model parameters in more
detail. The calculation of 
further observables reflecting different properties of the fireball expansion such as
the emission of electromagnetic probes or the suppression of high $p_t$ jets within the
same framework will help to confirm or disprove the outlined scenario.


\begin{acknowledgments}

I would like to thank S.~A.~Bass and B.~M\"{u}ller for helpful discussions, comments and their
support during the preparation of this paper.
This work was supported by the DOE grant DE-FG02-96ER40945 and a Feodor
Lynen Fellowship of the Alexander von Humboldt Foundation.
\end{acknowledgments}


\begin{thebibliography}{99}

\bibitem{Lattice}
F.~Karsch, E.~Laermann and A.~Peikert,
Phys.\ Lett.\  {\bf B 478} (2000) 447.

\bibitem{Synopsis}
T.~Renk,
hep-ph/0403239.


\bibitem{Brahms-dN-deta}
I.~G.~Bearden {\it et al.}  [BRAHMS Collaboration],
Phys.\ Rev.\ Lett.\  {\bf 88} (2002) 202301.

\bibitem{Brahms}
I.~G.~Bearden {\it et al.}  [BRAHMS Collaboration],
nucl-ex/0312023.

\bibitem{Quasiparticles}
R.~A.~Schneider and W.~Weise,
Phys.\ Rev.\ C {\bf 64} (2001) 055201;
M.~A.~Thaler, R.~A.~Schneider and W.~Weise,
hep-ph/0310251.

\bibitem{Hadrochemistry}
T.~Renk,
Phys.\ Rev.\ C {\bf 68} (2003) 064901.

\bibitem{ThermalPhenomenology}
E.~Schnedermann, J.~Sollfrank and U.~W.~Heinz,
Phys.\ Rev.\ C {\bf 48} (1993) 2462.

\bibitem{HBTReport}
U.~A.~Wiedemann and U.~W.~Heinz,
Phys.\ Rept.\  {\bf 319} (1999) 145.

\bibitem{HBTBoris}
B.~Tomasik and U.~A.~Wiedemann,
hep-ph/0210250.

\bibitem{Phenix-HBT}
S.~S.~Adler {\it et al.}  [PHENIX Collaboration],
nucl-ex/0401003.

\bibitem{Phenix-Spectra}
S.~S.~Adler {\it et al.}  [PHENIX Collaboration],
nucl-ex/0307022.

\end{thebibliography}
\end{document}